# Freely suspended, van-der-Waals bound organic nm-thin functional films: mechanical and electronic characterization

Lilian S. Schaffroth[1], Jakob Lenz[1], Veit Geigold[2], Maximilian Kögl[1], Achim Hartschuh[2,3,4], and R. Thomas Weitz[1,3,4*]

1: Physics of Nanosystems, Department of Physics, Ludwig-Maximilians-Universität München, Amalienstraße 54, 80799 Munich, Germany

2: Department of Chemistry, Butenandtstr. 5-13, 81377 Munich, Germany

3: Center for Nanoscience (CeNS), Schellingstraße 4, 80799 Munich, Germany

4: Nanosystems Initiative Munich (NIM), Schellingstraße 4, 80799 Munich, Germany

**email:** thomas.weitz@lmu.de

**Abstract:**

Determining the electronic properties of nanoscopic, low-dimensional materials free of external influences is key to discovery and understanding of new physical phenomena. An example is the suspension of graphene, which has allowed access to their intrinsic charge transport properties. Furthermore, suspending thin films enables their application as membranes, sensors, or resonators, as has been explored extensively. While the suspension of covalently-bound, electronically-active thin films is well established, semiconducting thin films composed of functional molecules only held together by van-der-Waals interactions could only be studied supported by a substrate. In the present work, it is shown that by utilizing a surface-crystallization method, electron conductive films with thicknesses of down to 6nm and planar chiral optical activity can be freely suspended across several hundreds of nm. The suspended membranes exhibit a Young's modulus of 2 to 13 GPa and are electronically decoupled from the environment, as established by temperature dependent field-effect transistor measurements.

Highly-ordered, molecularly-thin organic films are well known to the surface science community. They are typically manufactured via thermal evaporation of functional molecular constituents in ultra-high vacuum onto single-crystalline metallic surfaces and subsequently studied with respect to their local electronic or optical properties using scanning-tunneling methods[1]. While the established technologies allow routine control over the molecular arrangement with atomic precision[1], due to the strong interactions with the metal substrate, it is a big challenge to access the intrinsic electronic states of such molecular films. Consequently, current efforts are directed towards adding thin insulating layers between the conductive metal substrate and the organic film[2]. An alternative is the direct deposition of the functional molecules onto insulating substrates and subsequent study of their electronic behavior by lateral charge transport[3]. Here, the thickness of the films - which is usually on the order of tens of nms - and the presence of the insulating substrate pose the major challenges in addressing the intrinsic electronic properties[4, 5]. This inherent disorder might be one of the reasons why the fundamental charge transport mechanism in organic semiconducting thin films is still not fully understood[6]. Trapping mechanisms at the interface to the dielectric that deteriorate the electronic properties additionally limits the use of such films e.g. in sensors[7].

A known method to isolate electronic materials from a solid dielectric is implementing an air or vacuum interfacial layer (i.e. suspending the functional layer). For covalently bound materials, suspending nm thin films is an established method that has for example enabled access to the intrinsic electronic properties of graphene[8] and the mechanical behavior of 2D polymers[9]. Here, the ability to realize nm-thin membranes is attributed to the strong in-plane covalent bonds reflected, for example, in the extraordinarily high Young's modulus of graphene[10]. As van-der-Waals bonds are significantly weaker, the mechanical stability of purely van-der-Waals bound electronic layers such as that of organic thin film semiconductors is much more delicate. Therefore, the air-gap geometry has been successfully realized only for mm thick organic crystals[5]. However, detrimental bulk disorder could not be eliminated here[4], as air-gap gated semiconducting organic films of sufficiently low layer thickness have not been achieved up to now. A major step forward in this respect would therefore be the realization of a freely suspended, few-nm thin functional film based on purely van-der-Waals bound molecules.

With the goal to realize highly ordered organic electronically-active films that are on the one hand as thin as possible but on the other hand completely isolated from their environment, we have fabricated nanogap contact pillars as schematically sown in **Figure 1a**. In brief, we utilize standard electron-beam lithography, metal evaporation and lift-off techniques to pattern source/drain (S/D) contacts with a minimum separation of 100 nm. A brief wet-etch in hydrofluoric acid (HF) removes the $SiO_2$ around the contacts, and the S/D contacts remain as pillars that support the subsequently deposited semiconducting thin film. We have based the subsequent deposition of the electron-conductive organic semiconductor PDI1MPCN2 (N,N´-di((S)-1-methylpentyl)-1,7(6)-dicyano-perylene-3,4:9,10-bis(dicarboximide)) on our recently developed method of surface-mediated self-assembly at the solvent-air interface, which has shown to yield highly-crystalline, few nm-thin films[11]. The alkyl side-chains impose chirality on the molecules that can be observed in circular dichroism (CD) measurements (**Figure S1**). We have not been able to observe 3 nm thin suspended films corresponding to monolayer films, presumably due to their mechanical fragility or due to the inherent fluctuations present in 2D layers as predicted by the Mermin-Wagner theorem[12]. However, reproducible and targeted formation of 6 nm (i.e. bilayers) or thicker layers was realized on optimized (see methods) electrode structures (**Figure 1a** and **b**) with channel lengths $L$ between 100 nm and 450 nm and widths $W$ between 0.5 µm and 1 µm. Optical images capturing the deposition process that leads to the freely suspended organic semiconducting films are shown in **Figure 1**. . Atomic force microscopy (AFM) images of a 9 nm thin, freely suspended film is shown in **Figure 2a-d** and of a 6 nm thin film in **Figure S3** (please see the SI for a detailed discussion on the AFM height measurements).

To gain deeper understanding of the film's morphology, we have performed detailed cross polarized microscopy (**Figure S4**) and polarized confocal photoluminescence (PL) measurements (**Figure 2e**) on non-suspended thin films. Surprisingly, we have found that the chiral nature of the PDI1MPCN2 sidechains leads to a planar chirality of the thin film[13]: In these chiral thin films, the molecules change their in-plane orientation within a domain at an average in-plane twist angle between 1.2 and 0.5 °/µm. This change in orientation can be observed in the slowly varying PL intensity for linearly polarized excitation, which shows reversed contrast upon rotating the polarization by 90° (**Figure 2f,** and **Figures S5** and **S6**). Aside from thin films, we also find bulky structures on our samples, which are, upon closer inspection, composed of chiral nanowires (**Figure 2e** inset, more details in **Figure S1**). We will address the implications of the chiral nature of the films below.

The ability for a material to withstand free suspension critically depends on its mechanical strength that is in thin films mostly determined by the in-plane Young's modulus. At first sight, given the fact that the van-der-Waals force is significantly smaller than the covalent binding strength, it seems surprising that we are able to realize freely suspended thin films of only two molecules in thickness. However, liquid crystalline materials have been realized in free-standing from[14], meaning that in general the breaking strength of van-der-Waals bound materials is high enough. On the other hand, the functionality of liquid crystalline films does not depend on a sizable transfer integral between adjacent molecules, and it is a priori unclear if a charge transfer integral large enough to enable electronic transport between adjacent molecules can be preserved in suspended films. To the best of our knowledge, free-standing electronically-active van-der-Waals bound thin films have not been realized up to now.

To measure the structural integrity, we have assessed the mechanical properties of the films by measuring the in-plane Young's modulus of suspended layers via AFM-based nanoindentation. Given that the calculation of the Young's modulus in the geometry realized in **Figure 2** is challenging since it cannot be determined from measurements on a single device, we have suspended films overcircular cavities in thin gold pads (**Figure 3a,** see also **Figure S7-S9** for more experimental details). Representative force-indentation curves are shown in **Figure 3b** (see **Figure S10** for further measurements).

The 3D Young's modulus of the thinnest film (~ 10 nm) we have measured is 2 GPa, while for the thickest film of 18 nm the Young's modulus was determined to 13 GPa. These values are remarkably high, given that the films are only few molecular layers thin, have been crystallized from solution and are only held together by van-der-Waals interactions. In comparison, a Young's modulus of 14 GPa has been reported for free-standing covalently bound 2D polymers[9] and a Young's modulus of 5 – 15 GPa was measured for > 25 nm thick organic semiconductor films on a substrate support[15]. Graphene, for comparison, has a perfectly crystalline, covalently bound lattice and consequently a much higher Young's modulus of 1TPa[16]. Considering that the van-der-Waals force is at least two orders of magnitude weaker than the covalent bond, our determined values for the Young's modulus are a clear indication of the high crystal quality of our suspended chiral organic thin films.

Besides the mechanical properties of the suspended thin films, we have also verified that they are electronically decoupled from their environment. For these measurements, we have used the electrode structures introduced in **Figures 1** and **2**. The gate dielectric of these structures is composed of the air-gap and the residual $SiO_2$ layer. Just as for FETs based on non-suspended PDI1MPCN2-thin films[11], characteristic electron transport behavior (**Figure 4a and b**) with a well-defined subthreshold swing as well as a threshold value close to $V_{GS} = 0$ V can be measured. Furthermore, the devices have a negligible hysteresis when measured in vacuum, which stems from an absence of fixed charges as can be expected, since the air-gap dielectric does not allow for trapping of charges directly at the conductive channel. At a charge carrier density of n = 9 x $10^{10}$ 1/cm², a charge carrier mobility of $\mu_{lin} = 0.4 \cdot 10^{-3}$ cm² / Vs was determined from the linear regime of a transfer curve measured at $V_{DS}$ = 1 V. A slightly higher mobility of $\mu_{sat}$ = 0.6·10$^{-3}$ cm$^2$ / Vs was extracted from the saturation regime of a transfer curve measured at $V_{DS}$ = 8 V. The corresponding output curve is shown in **Figure 4c**.

Having realized highly-ordered, freely-suspended semiconducting thin films for the first time gives us the unique chance to explore the temperature dependency of charge transport free from detrimental influences both from a solid dielectric[5, 17] and from within the bulk of the semiconductor [4, 18]. Typical temperature dependent measurements of charge carrier mobility as well as of the threshold voltage and the interface state trap density $N_{it}$, as evaluated from the subthreshold slope, can be seen in **Figure 4d-f**. Representative, temperature dependent transfer curves are shown in **Figure S11** in the SI. Mean $N_{it}$ and $V_{th}$ values were calculated as an average over three comparable devices measured under the same conditions, each based on a suspended bilayer (6 nm thick). The temperature dependent measurements show that the mobility freezes out at low temperatures, possibly due to contact effects. In the high temperature regime however, we find that unlike to 3 to 9 nm thin organic films located on a substrate (where disorder from the non-current carrying part of the semiconductor already was eliminated due to the thin channel)[17], the charge carrier mobility does not decrease but continues to increase, consistent with a decoupling of the film from the dielectric. The difference in temperature evolution might however also be related to the different charge carrier densities of the two measurements: the mobilities in the thin suspended films were evaluated at a two orders lower charge carrier density than the measurements performed in the substrate-supported films. Therefore, a measurement that is more density independent would be beneficial. We consequently have also evaluated $V_{th}$ and $N_{it}$, that show a monotonous behavior as function of temperature. This is unlike to previous measurements of the same semiconductor when located on a substrate, where we recently identified traps at the semiconductor/dielectric interface as limiting factor for charge transport leading to an increased $V_{th}$ and $N_{it}$ towards elevated temperatures[17] (see **Figures 4d** and **e**).

Furthermore, the absolute $N_{it}$ -values of the horizontally suspended layers (in the order of $10^{12}$ - $10^{13}$ 1/(cm$^2$ V)) are consistently more than one order of magnitude smaller than those of the non-suspended layers ($10^{13}$-$10^{14}$ 1/(cm$^2$ V)), reflecting the suppression of trapping at the semiconductor/dielectric interface in the suspended films. To ensure that the temperature changes are not detrimental to the electronic properties of the suspended films, after performing the measurement at the maximum temperature of 450 K we have directly cooled the device to 100 K, repeated the temperature cycle, and found that the devices were still functional.

Our measurements reveal that, by spatially separating the semiconducting layer from the solid dielectric by vacuum, scattering of charge carriers with the dielectric can be avoided. The results also indicate that charge transport in FETs based on non-suspended thin films is significantly impacted by scattering at the dielectric at temperatures above 250 K.

While the temperature dependence of *µ*, $V_{th}$ and $N_{it}$ above 250 K show that in the absence of a solid dielectric scattering is suppressed, the absolute charge carrier mobility of the suspended films is 2-3 orders of magnitude lower than that of non-suspended films. While we can currently only speculate about the microscopic reason, it might be related to strain, folds or bends in the organic film[17] or to contact effects (see **Figure S12** for an output curve at 150 K revealing injection problems) since the channels are only several 100 nm long and we are operating the device in a bottom-gate bottom-contact geometry[19]. In an effort to exclude the impact of contacts, we have realized suspended films with four terminals as shown in **Figure S13** to allow for a measurement of the electrical properties using the gated van-der-Pauw method[20]. In these four-terminal measurements the mobility was found to be less than a factor of seven larger compared with the two-terminal measurements, indicating that the contacts only partially limit carrier injection.

**Summary and conclusion**

We have realized freely-suspended, two molecule thin planar chiral organic semiconducting films free from detrimental impact from a solid dielectric support for the first time. The films are composed of conjugated perylene diimide small molecules and held together by weak van-der-Waals forces. Temperature-dependent charge transport measurements confirm that the thin films are indeed decoupled from their environment. The large Young's moduli of up to 13 GPa observed for freely suspended films clearly demonstrate their excellent crystal quality. We anticipate that the here developed technique can provide a general approach to the suspension of functional van-der-Waals bound small molecule films and to the investigation of their charge transport behavior free from environmental impacts.

Remarkably, our deposition method can be used to translate the chiral structure of the molecules into a chiral ordering within the freely-suspended film which gives inspiration for further studies on the chiral nature of the PDI1MPCN2 molecules and the in-plane chirality of the PDI1MPCN2 films. For example, potential coupling between charge carrier motion and the chiral nature of the film[21] could be potentially investigated in these suspended films. Finally, the aspects on in-plane chirality[13] – the first observation of its kind in such semiconducting films to the best of our knowledge – will possibly allow use of this new class of organic films in novel optoelectronic devices.

## Materials and Methods

### Fabrication of electrode structures

We have used highly doped Si wafers coated with 300 nm of thermally grown $SiO_2$. Standard electron beam lithography and development were used to pattern the PMMA resist. 80nm of gold with a 1nm thick chromium adhesive layer were subsequently deposited using ultra-high vacuum electron beam evaporation. A sufficient amount (50nm-150nm) of the silicon oxide surface layer was wet-etched in hydrofluoric acid (HF) to ensure that the slightly bent few layer films are not in contact with the substrate surface.

**Crystallization of the suspended films**

The high-surface tension, high viscosity solvent dimethyl phtalate (DMP) was selected to enable surface-mediated growth at the liquid-air interface and to minimize perturbation of the crystalline molecular organization induced by Marangoni flow. The mixing ratio between molecule and solvent of 0.1wt% in combination with the solvent volume of 2-3µl provides continuous thin film growth and at the same time minimal formation of parasitic, large, disordered agglomerations. The growth temperature of 70 °C is optimized with respect to grain size and crystallization duration. As temperature decreases, crystallization increasingly sets in at nuclei spontaneously forming on the droplet surface wherefrom it continues radially outwards forming 2-dimensional circular structures referred to as spherulites. Since an increase in spherulite density is accompanied by a continuous decrease of the average grain size, the spherulitic growth mode is strategically avoided by the choice of temperature. Increasing the growth temperature on the other hand impedes the molecular alignment leading to a decrease in molecular order and grain size.

A critical point in realizing the formation of suspended layers is targeted multilayer formation that can be induced by drastically reducing the droplet volume as well as by employing the contact design introduced in the main part that allows for the accumulation of molecules at a designated position. By confining the droplet to the immediate contact area, which was achieved by brushing the solvent across the contacts with nitrogen at a low pressure directly after drop-casting (**Figure 1a**), the formation of thick, amorphous agglomerates at the droplet edge, that serve as a preferable crystallization site for residual molecules, can be avoided.

The adhesive forces between solvent and gold-coated contact pillars cause the liquid to wet the structures' edges, while the cohesion of the solvent ensures the extension of the droplet between the source and drain contact structures.

As the solvent evaporates, it is drawn along the triangular notches towards the narrow channel at the electrodes' center, where the molecules accumulate and few-layers form (**Figure 1b-d**). Suspension of the few-layer films is achieved if the pinning forces between thin film and the gold contact as well as the intermolecular forces within the film dominate the adhesive forces between thin film and solvent as it evaporates (see **Figure S2** for a schematic of the anticipated forces acting during the drying process). An example of a resulting suspended thin film can be seen in **Figure 1**.

**Nanoindentation measurements**

Gold pads with off-centered circular cavities (typical diameters used for indentation measurements: 200-400 nm) were realized using e-beam lithography as described above. To allow for the solvent to escape from the thin-film covered cavities during the deposition process and air during an indentation measurement, gold pads were underetched until one side of the cavity was exposed to ambient (**Figure 3a** reveals a tear in the organic film that allows the solvent to escape from the cavity).

All nanoindentation measurements were performed with a Dimension 3100 AFM and cantilevers (k=40N/m) with a monolithic silicon probe (d<10nm). Details can be found in the SI.

**Electrical measurements**

All measurements were performed in a LakeShore cryogenic probestation in high-vacuum. The widths and lengths of the channels were determined from AFM images after performing the electrical measurements. To allow for good comparability of the interface state density of the different samples at each temperature level, the logarithmic subthreshold slope was determined using the following procedure: The derivative of the smoothed drain-current in logarithmic application with respect to $V_{GS}$ was calculated, and the mean $d\log(I_D)/dV_{GS}$ determined in a gate voltage interval of ±0.8V around the maximum.

The 4-terminal gated van-der-Pauw method was used to approximate the impact of contact resistance as it allows for a calculation of sheet resistance for thin films of arbitrary shape[20]. Further details can be found in the SI.

## Acknowledgements

We acknowledge funding from the Center of Nanoscience (CeNS), the Nanosystems Initiative Munich (NIM) and the initiative "Solar technologies go Hybrid" (SolTech). We also acknowledge help from T. Liedl and T. Funck for help with the CD measurements. We acknowledge BASF SE for providing the organic semiconductor.

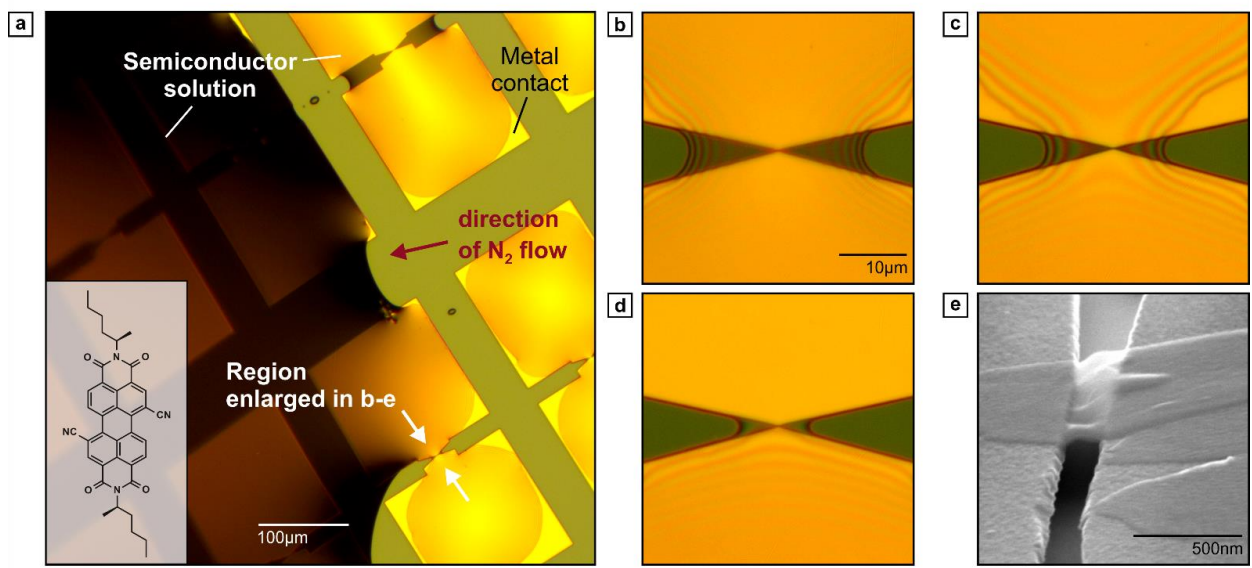

**Figure 1: Images capturing the deposition process of the suspended organic thin films. a)** Top view of the substrate as the semiconductor solution is N₂-brushed over the gold contacts. **b)** to **d)** As the solution evaporates, it is drawn towards the trench at the position of smallest separation between the contact pairs, where the molecules accumulate and multilayer formation is initiated. **e.)** SEM image of the final device showing the organic film freely suspended between the two metal contacts.

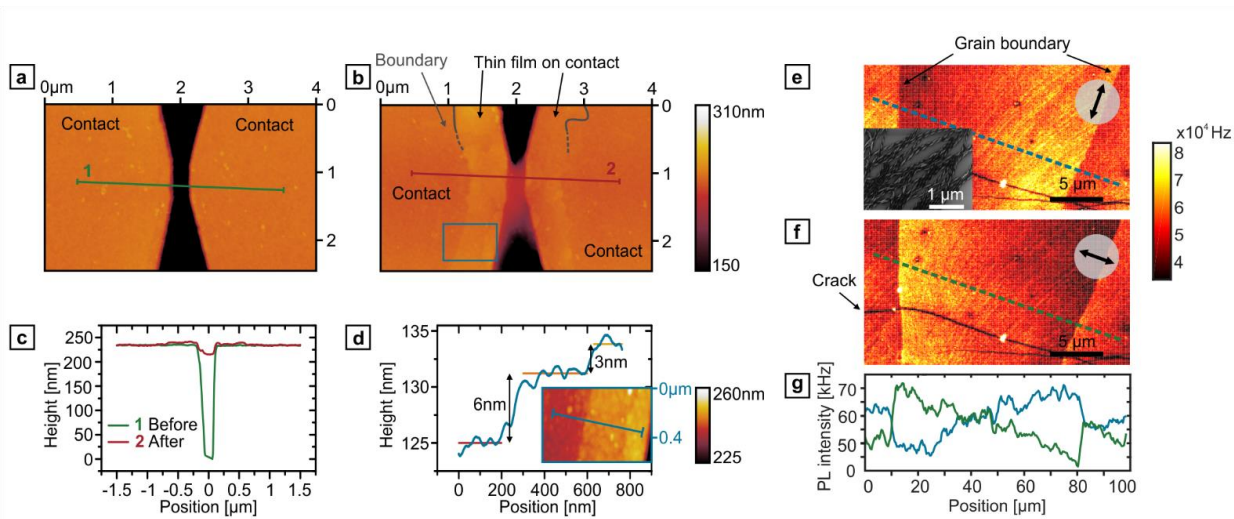

**Figure 2: Atomic force microscopy (AFM) of freely suspended organic thin film and optical characterization of non-suspended films.** AFM image of a nanogap device prior **a)** and after **b)** deposition of a thin organic film. **c)** Line-profiles along the gap showing the suspension and slight bending of the deposited thin film. **d)** Detail of the outlined area in **b)** (inset) and line-profile revealing that the organic film is 9 nm thin. **e)** and **f)** Confocal photoluminescence (PL) microscopy images of a one-molecule thin film on a glass substrate detected upon linearly polarized laser excitation as indicated by the arrows. The observed PL intensity variations are due to a spatially varying orientation of the in-plane component of the transition dipole moment as can be seen from the contrast reversal after 90° rotation of the excitation polarization. The gradual PL intensity variations indicate a continuous rotation of the molecular orientation within domains on a length scale of several tens of micrometers. Domain boundaries with sudden changes in molecular orientation are seen as steps in the PL intensity profiles. Inset to **e)** SEM image of chiral nanowires of the same molecule. **g)** Extracted line-profiles of the PL intensity from **e)** and **f)**.

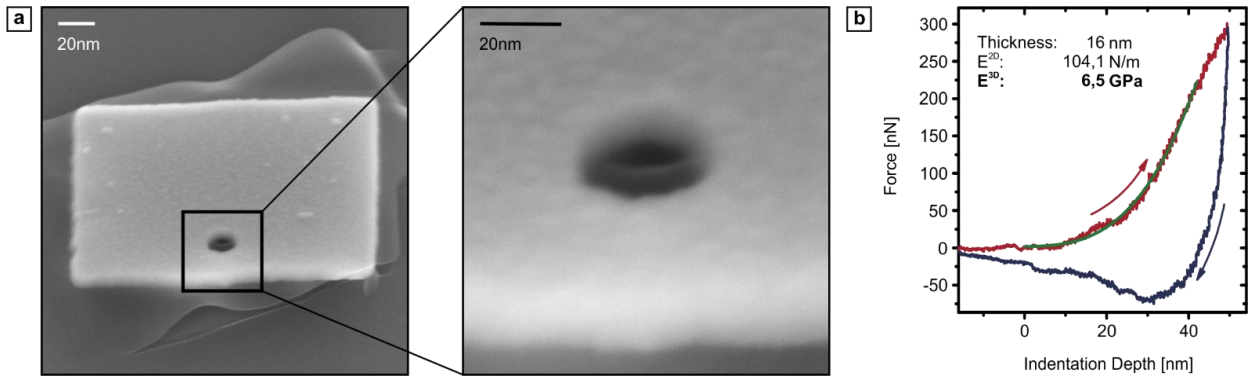

**Figure 3: Nanoindendentation measurements of suspended thin films. a)** SEM images of a thin film suspended over a circular cavity in a gold pad fabricated by electron beam lithography. **b)** Force-indentation characteristic of a 16 nm thin organic film.

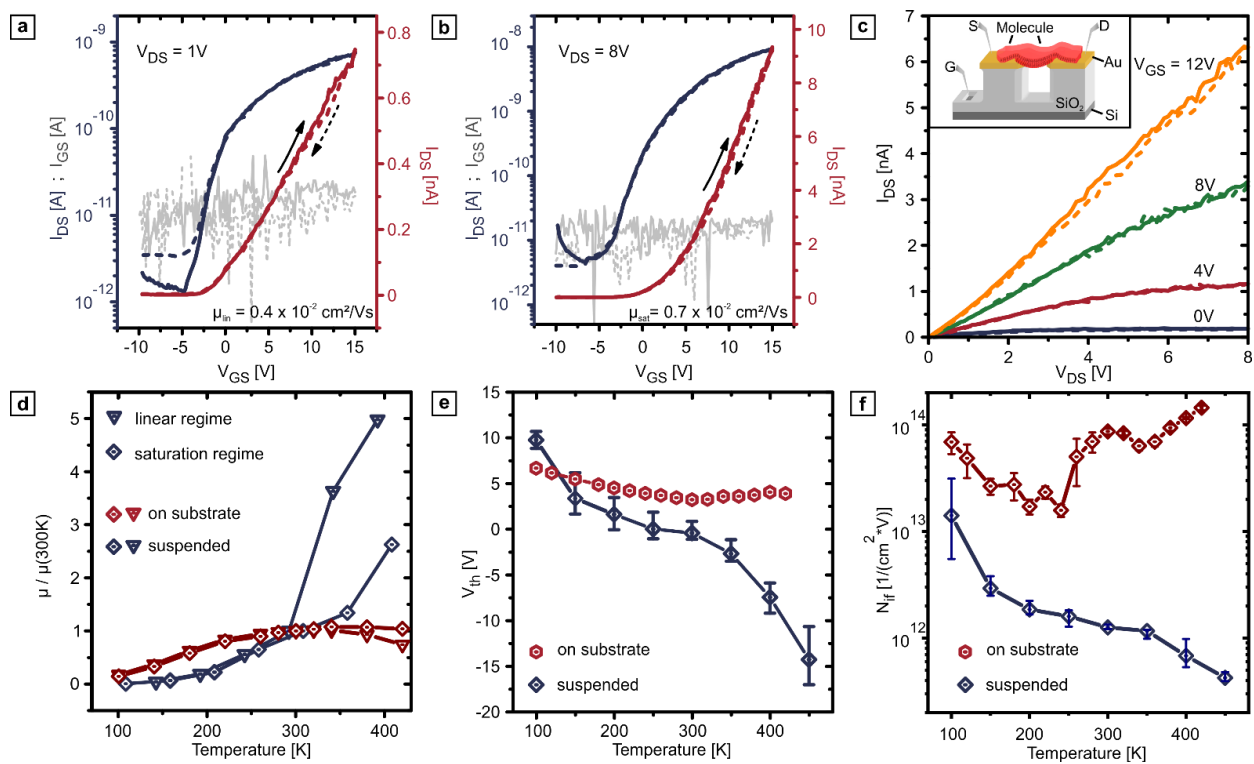

**Figure 4: Electrical characterization of suspended, few molecule thin organic films. a)** and **b)** Transfer curves of a suspended organic film (6 nm thin) and determined mobilities in the linear and saturation regime. **c)** Corresponding output curve. Inset to **c)** Schematic of a device based on a suspended bilayer and the electrical contacting of gate (G), source (S) and drain (D). Probe needles were utilized to measure the electrical characteristics of the device. **d)** Temperature dependence of the normalized linear and saturation mobility of a suspended semiconducting film compared with measurements of a film directly deposited onto a Si+30nm $Al_2O_3$ substrate support. **e)** Temperature dependence of the threshold voltage and **f)** of the interface state density averaged over three freely suspended films compared to the data obtained for films on substrate supports. The bars indicate the minimum and maximum values measured for the individual devices. The data for non-suspended films shown in **d)** - **f)** was taken from Vladimirov et. al [17].